\documentclass{svjour3}       
\smartqed  
\usepackage[utf8]{inputenc}
\usepackage[T1]{fontenc}
\usepackage[colorlinks = true,
            linkcolor = blue,
            urlcolor  = blue,
            citecolor = blue,
            anchorcolor = blue]{hyperref}
\usepackage{multirow}%
\usepackage{amsmath,amssymb,amsfonts}%
\usepackage{mathrsfs}%
\usepackage[title]{appendix}%
\usepackage{textcomp}%
\usepackage{manyfoot}%
\usepackage{braket}
\usepackage{dsfont}
\usepackage{amsmath} 
\usepackage[numbers, sort & compress]{natbib} 
\usepackage{textcomp}
\usepackage[utf8]{inputenc}
\usepackage{graphicx}
\usepackage{dcolumn}
\usepackage{bm}
\usepackage{xcolor}
\usepackage{physics}
\usepackage{float}
\usepackage{algorithm}
\usepackage{caption}
\usepackage{subcaption}
\usepackage{hyperref}
\begin{document}

\title{Machine Learning assisted noise classification
with Quantum Key Distribution protocols.}

\author{Shreya Banerjee       \and
        Ashmi A \and
        Prasanta K. Panigrahi}

\institute{Shreya Banerjee \at
           Center for Quantum Science and Technology, Siksha 'O' Anusandhan University, 
         \email{shreya93ban@gmail.com}
          \and
Ashmi A. 
\at Department of Physics, Indian Institute of Science Education and Research Pune, Pune-411008, Maharashtra, India,
\and
          Prasanta K. Panigrahi \at
Center for Quantum Science and Technology, Siksha 'O' Anusandhan University, Bhubaneswar, 751030, India,
\at Indian Institute of Science Education and Research Kolkata, Mohanpur- 741246, West Bengal, India,
          \email{director.cqst@soa.ac.in}
}

\maketitle

\begin{abstract}
{We propose a hybrid protocol to classify quantum noises using supervised classical machine learning models and simple quantum key distribution protocols. We consider the quantum bit error rates (QBERs) generated in QKD schemes under consideration of different noises, and identify the noise channels with high accuracy for both training and test data. Our protocol classifies quantum noises with high accuracy under assumption of two different scenarios; in one case we assume two remotely located parties share keys through noisy quantum channels, whereas, in the second case, we simulate the QKD protocols on a gate-based quantum computer, where the gates are afflicted with noise. Alongside efficient classification, our work also throws light on the difference in distribution characteristics of QBERs generated in these two scenarios. Finally, our method is based on classical post processing of data generated from very simplistic quantum protocols, making it readily implementable in the current era of noisy quantum computing with low number of qubits.   }
\end{abstract}

\keywords{Quantum Clustering, Unsharp Measurements, Quantum Algorithms}

\section{Introduction} 
With the recent hardware developments the widely discussed field of quantum information and computation is now at a crossroad, where real world applications are only possible with several rounds of error correction and/or mitigation \cite{willow2024, harddev2, harddev3, hardev3}. Due to limitations in scalability and noisiness of quantum hardware, deployment of error correction schemes on quantum algorithms and protocols is still at infancy \cite{willow2024, qec1, qec2}. However, error mitigation schemes based on classical post-processing offers a timely and viable solution to this problem \cite{Qem1, QEM2, QEM3, QEM4, QEM5}. Many of these error mitigation schemes work better for specific noises, and thus, characterization of quantum noises remains a topic of avid interest \cite{PhysRevA.109.042620}. There have been several approaches to classify quantum noises \cite{Paschottaquantum_noise, Mukherjee_2024, Martina_2023}, with examples that use  machine learning techniques. As examples, in \cite{Mukherjee_2024}, for classifying quantum and classical parameters of dephasing noise neural networks were employed. Neural networks were also used in \cite{Martina_2023} to classify Markovian noise and three different classes of non-Markovian noise.

Among applications of quantum principles in quantum technology, secure communication is a prime example \cite{qcomm,qcomm1,qcomm2}, that has seen real-world deployment \cite{rqcomm1, rqcomm2}. It relies on the simplest forms of quantum correlations, such as, quantum superposition and the property of quantum measurement leading to collapse of two wavefunction. Its practical application necessitates its exposure to quantum decoherence arising from environmental effects apart from the error arising from state preparation and measurement.  
The fact that the crucial step of quantum secure communication relies on generating a secure key comprising of binary numbers, offers the possibility of modeling and observing various types of errors. The role of non-quantum pre and post processing operations also observe the possibility of error detection and mitigation in a hybrid model. Two or multiparty quantum communication requires generation and distribution of a quantum key, known as quantum key distribution.  

Quantum key distribution (QKD) protocols \cite{BENNETT20147,Ekert1992,PhysRevLett.68.557} are used to securely distribute encryption keys between two or more participants. BB84 \cite{BENNETT20147} and BBM92 \cite{PhysRevLett.68.557} protocols are examples of two very simple QKD schemes, which only requires single and two-qubit operations whether simulated on a gate-based quantum computer \cite{qiskit2024} or in a more general photonic setting using quantum channels\cite{}. When simulated with noisy operations, it often leads to quantum bit-error in a generated encryption key.  In this work, we show that from the distribution of quantum bit error rate (QBER) of these simple QKD schemes, it is possible to identify the noises present in quantum channels (or gates in a gqte-based quantum computer) by using supervised machine learning algorithms such as K-Nearest Neighbor \cite{KNN1,KNN2,KNN3}, Gaussian Naive Bayes \cite{GNB1,GNB2,GNB3} and Support Vector Machines with simple kernels \cite{SVM1,SVM2,SVM3,ker,rbf}. Our results show high accuracy for binary classification of amplitude damping, bit-flip and depolarizing noise models. The key innovation of  our approach lie at its simplicity, as we use very simple circuits, and perform basic statistical analysis, and employ low-depth classification algorithms.

We explain the QKD protocols, and noise channels used in this work alongwith the quantum bit error rate (QBER) in Sec.~\ref{prereq}. In Sec.~\ref{method}, we describe our methodology for classification of quantum noises, and present our results for classifying noisy quantum channels. In Sec.~\ref{qiskit_results}, we show the applicability of our method on a gate-based quantum computer, and finally summarize our findings in Sec.~\ref{conc} with future directions.   

\section{Prerequisites}\label{prereq}
Our proposed protocol deals with two different quantum key distribution protocols, under assumption of three different noise channels in different scenarios. We use Quantum Bit Error Rate (QBER) as our metric to determine the effect of noises in a QKD scheme. We now briefly explain these key concepts.\\
\vspace{1 mm}

\noindent\textbf{QKD Protocols:} We have chosen two very simplistic QKD protocols for our classification method, namely, BB84 \cite{BENNETT20147} and BBM92 \cite{PhysRevLett.68.557}. 
\vspace{1 mm}

\noindent\textit{BB84 protocol:} We consider two parties, Alice and Bob who participate in a quantum key distribution scheme using BB84 protocol \cite{BENNETT20147}. Alice chooses a $l$-bit long input message, and prepare $l$ qubits in quantum states according to the input strings. Further, Alice and Bob each selects a sequence of $l$ measurement bases, randomly chosen from computational and diagonal bases. Alice measures her qubits according to her sequence of measurement bases and sends them to Bob. Bob measures them according to his sequence of bases. Next, they publicly announce their basis-sequence, and only keep the outcomes where both Alice and Bob have same bases. Ideally in a noiseless scenario, where there is no eavesdropper, Alice and Bob will get a key with length $L \leq l$. 
\vspace{1 mm}

\noindent \textit{BBM92 Protocol:} BBM92 protocol \cite{PhysRevLett.68.557} is an entanglement based QKD protocol, where Alice and Bob share l pairs of entangled photon (Bell State) Similar to BB84, Alice and Bob chose their random bases independently and measure according to their own basis sequences. . Afterwards, they declare their bases, and discard the outcomes where they did not choose the same basis. In an idealistic scenario, they retrieve the same key with length $L \leq l$.\\
\vspace{1 mm}

\noindent\textbf{Quantum Noise Channels:}
 To check the effect of quantum noise on QKD schemes, we have considered three different noise channels: amplitude damping noise, bit-flip noise, and depolarizing noise. For our analyses, we have used the operator-sum representation of quantum noise channels that models the impact of noises on a quantum state $\rho$ as follows, 
\begin{equation}\label{opsum}
\mathcal{E}(\rho)= \sum_{k} E_{k}\rho E_{k}^{\dag},
\end{equation}
where, operators $E_{k}$ are the operational elements satisfying $\sum_{k}E_{k} E_{k}^{\dag} = \mathds{I}$. The complete set of operators $\{ E_k \}$ provides a way of analyzing different quantum noises in the Hilbert space of the quantum state itself.\\
\noindent The \textit{bit-flip noise channel} flips the state of a qubit from $\ket{0} $ to $\ket{1}$ (and vice-versa) with probability $p$, where $p$ is the strength parameter. The operational elements, also known as Kraus operators are provided as \cite{Nielsen_Chuang_2010},
\begin{equation}
   E_0 = \sqrt{1-p} \mathds{I}; \hspace{2 mm} E_1 = \sqrt{p} \mathds{\sigma}_{x},
\end{equation} 
here $\mathds{\sigma}_{x}$ is the Pauli-x matrix.

\noindent The \textit{amplitude damping noise channel} models  the energy dissipation in a quantum channel. The operational elements are provided as \cite{Nielsen_Chuang_2010},
\begin{equation}
   E_0 = \left( \begin{array}{cc} 1 & 0 \\ 0 & \sqrt{1-p}  \end{array}\right); \hspace{2 mm} E_1 = \left( \begin{array}{cc} 0 & \sqrt{p} \\ 0 & 0  \end{array}\right),
\end{equation} 
here $p$ is the strength of the noise.

\noindent The \textit{depolarizing noise} can be seen as a map that takes one quantum state $\rho$ to a linear combination of itself and the maximally mixed state $\mathds{I}$, based on the strength parameter $p$ \cite{Nielsen_Chuang_2010}. The operational elements for this noise channels are provided as, 

\begin{equation}\label{krauss_dp}
   E_0 = \sqrt{1-\frac{3p}{4}} \mathds{I}; \hspace{2 mm} E_1 = \sqrt{\frac{p}{4}}\mathds{\sigma}_{x}; \hspace{2 mm}
   E_2 = \sqrt{\frac{p}{4}}\mathds{\sigma}_{y} \hspace{2 mm} E_3 = \sqrt{\frac{p}{4}}\mathds{\sigma}_{z},
\end{equation} 
here $\mathds{\sigma}_{x, y, z}$ are the Pauli matrices. \\
\vspace{2 mm}

\noindent\textbf{Quantum Bit Error Rate (QBER):} We consider that in the quantum channel used by Alice and Bob for QKD protocols are afflicted with noise. Then the key generated at Alice's end will differ from key generated at Bob's end. The metric to encapsulate this error is called Quantum bit error rate (QBER), and is given by,
\begin{equation}\label{eq_qber}
    \text{QBER} = \frac{\text{Number of mismatched key bits}}{\text{Number of same measurement bases for Alice and Bob}}.
\end{equation}
In this work, we use this metric to compute the effect of different noise channels on the QKD schemes in different scenarios. \\
\vspace{2 mm}

\section{Protocol for quantum noise classification with QKD schemes:}\label{method}
We assume that Alice and Bob share a private key through a simple quantum key distribution scheme, where Alice prepares and transfers qubits to a remotely located Bob through a quantum channel. Under the assumption of an open system, the quantum channel interacts with the environment, and becomes noisy. In this work, we have considered two distinct quantum noises that can affect quantum channels: amplitude damping and the bit-flip noise, and shown that it is possible to distinguish the noisy quantum channels with high accuracy from the quantum bit error rates generated in a QKD scheme. 

The entire classification process can be broadly divided into three segments: 
\begin{itemize}
    \item Performing a simple QKD scheme on a quantum noise channel and accumulating the Quantum Bit Error Rates (QBERs) by comparing Alice and Bob's outcome keys.
    \item Statistical analysis of accumulated QBERs to prepare a dataset with features relevant to identifying the noisy channel.  
    \item Implementing classical machine learning algorithms to identify the nature of the noise. 
\end{itemize}
We now explain these steps in detail with BB84 protocol as an example of a simplistic QKD scheme. \\
\vspace{5mm}

\noindent\textbf{Simulation of QKD scheme under assumption of Noise:}
\begin{figure}[hbt!]
    \centering
    \includegraphics[scale=0.35, trim={2cm, 5cm, 2cm, 3cm},  clip]{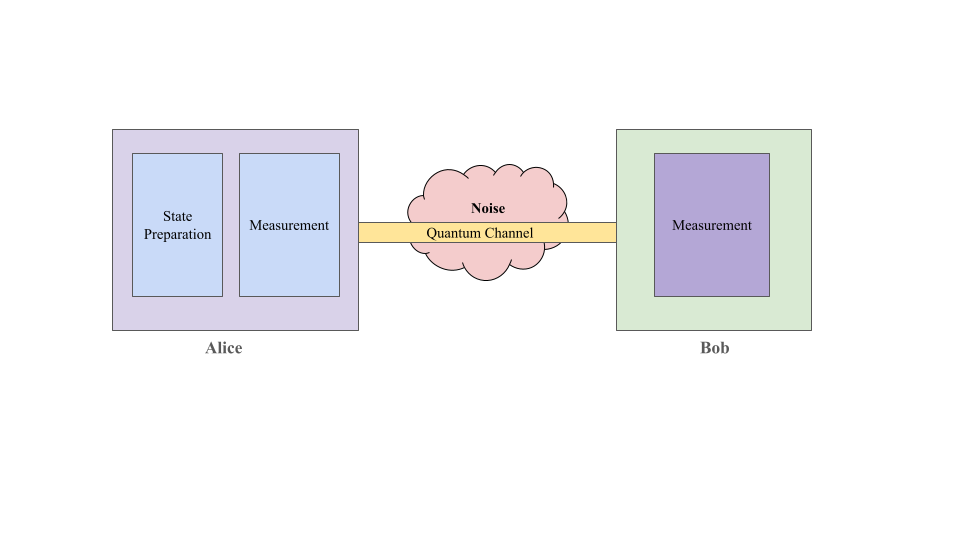}
    \caption{Schematic of BB84 QKD scheme under the assumption of noisy quantum channel.}
    \label{noisy_channel_qkd}
\end{figure}

We assume that Alice and Bob use BB84 protocol to generate their respective keys. As shown in Fig~\ref{noisy_channel_qkd}, we considered the noise impacts the quantum channel during the transfer of qubits from Alice to Bob, and there is no noise present during either state preparation or measurement. Under these assumptions, we simulated the BB84 scheme in presence of amplitude damping and bit-flip noises. We have used the operator-sum representation for both the noise channels, as explained in Sec.~\ref{prereq}, and computed the quantum state under amplitude damping and bit-flip noises at each step of the algorithm, i.e., after Alice sends her qubit to Bob, and after Bob measures the state based on his chosen basis. Finally, we associated the probability of Bob receiving bit-string '0' or '1' with the diagonal elements of the quantum state after Bob's measurement.

\begin{table}[hbt!]
    \centering
    \begin{tabular}{|c|c|c|c|c|c|c|}
        \hline
        
        Bit & Alice's basis & $\rho$ & $\mathcal{E}(\rho)$ & Bob's basis & RP('0')  & RP('1') \\
        \hline
         '0' & Computational &  
         $\left( \begin{array}{cc} 1 & 0  \\ 0 & 0 \end{array}\right)$ & $\left( \begin{array}{cc} 1-p & 0 \\ 0 & p  \end{array}\right)$
 & Computational & $1-p$  & $p$   \\

          \hline
         '1' & Computational & 
         $\left( \begin{array}{cc} 0 & 0  \\ 0 & 1 \end{array}\right)$ 
         & $\left( \begin{array}{cc} p & 0 \\ 0 & 1-p  \end{array}\right)$ 
         & Computational & $p$  & $1-p$   \\

           \hline
         '0' & Diagonal & 
         $\frac{1}{2}\left( \begin{array}{cc} 1 & 1  \\ 1 & 1 \end{array}\right)$ 
         & $\frac{1}{2}\left( \begin{array}{cc} 1 & 1  \\ 1 & 1 \end{array}\right)$ 
         & Diagonal & $1$  & $0$   \\

     \hline
         '1' & Diagonal & 
         $\frac{1}{2}\left( \begin{array}{cc} 1 & -1  \\ -1 & 1 \end{array}\right)$ 
         & $\frac{1}{2}\left( \begin{array}{cc} 1 & -1  \\ -1 & 1 \end{array}\right)$ 
         & Diagonal & $0$  & $1$   \\

    \hline
    
    \end{tabular}
    \vspace{2 mm}
    \caption{Probability of erroneous bits generated by Bob after key generation through BB84 protocol with a noisy (bit-flip) quantum channel. Here, 'bit' refers to the key bit encoded by Alice. $\rho$ and $\mathcal{E{(\rho)}}$ represent quantum state transferred by Alice, and noisy state received by Bob respectively. RP('0') and RP('1') provide Bob's retrieval probability of the key bit as '0' and '1'. Only the results for same measurement bases used by Alice and Bob are provided here.}
    \label{tab_qstates_bp}
\end{table}

\begin{table}[hbt!]
    \centering
    \begin{tabular}{|c|c|c|c|c|c|c|}
        \hline
        
        Bit & MB & $\rho$ & $\mathcal{E}(\rho)$ & RP('0')  & RP('1') \\
        \hline
         '0' & C &  
         $\ket{0}\bra{0}$ 
         
         & $\left( \begin{array}{cc} 0.5 & 0 \\ 0 & 0.5  \end{array}\right)$
  & $0.5$  & $0.5$   \\

          \hline
         '1' & C & 
         $\ket{1}\bra{1}$ 
         & $\left( \begin{array}{cc} p & 0 \\ 0 & 1-p  \end{array}\right)$ 
         & $p$  & $1-p$   \\

           \hline
         '0' & D &
         
           $\ket{+}\bra{+}$
         
         & $\frac{1}{3}\left( \begin{array}{cc} 1+p & \sqrt{p} + \sqrt{1-p}  \\ \sqrt{p} + \sqrt{1-p} & 2-p \end{array}\right)$ 
         
          & $\frac{1}{2}+\frac{\sqrt{p}+\sqrt{1-p}}{3}$  & $\frac{1}{2}-\frac{\sqrt{p}+\sqrt{1-p}}{3}$   \\

     \hline
         '1' & D & 
         
          $\ket{-}\bra{-}$ 
         
         &  $\frac{1}{3}\left( \begin{array}{cc} 1+p & -(\sqrt{p} + \sqrt{1-p})  \\ -(\sqrt{p} + \sqrt{1-p}) & 2-p \end{array}\right)$

          & $\frac{1}{2}-\frac{\sqrt{p}+\sqrt{1-p}}{3}$  & $\frac{1}{2}+\frac{\sqrt{p}+\sqrt{1-p}}{3}$   \\

    \hline
    
    \end{tabular}
    \vspace{2 mm}
    \caption{Probability of erroneous bits generated by Bob after key generation through BB84 protocol with a noisy (amplitude damping) quantum channel. Here, 'Bit' refers to the key bit encoded by Alice. $\rho$ and $\mathcal{E{(\rho)}}$ represent quantum state transferred by Alice, and noisy state received by Bob respectively. RP('0') and RP('1') provide Bob's retrieval probability of the key bit as '0' and '1'. Only the results for same measurement bases (mentioned as 'MB') used by Alice and Bob are provided here.}
    \label{tab_qstates_ad}
\end{table}

Tables~\ref{tab_qstates_bp} and \ref{tab_qstates_ad} explain the theoretical workflow of the noisy key simulation process of BB84 protocol under bit-flip and amplitude damping noises respectively. The tables present how the noises affect the the quantum state $\rho$ sent by Alice to Bob, who receives the state $\mathcal{E}(\rho)$. Accordingly, the retrieval probability of the original key bit changes with certain choices of Alice and Bob's measurement bases. As an example, when Alice decides to send key bit '0', and measures it in the diagonal basis, prior to transfer, the qubit is at $\ket{+}$. However, if the channel is noisy under amplitude damping noise, it can be seen from Table~\ref{tab_qstates_ad}, the state Bob receives is in state $\frac{1}{3}\left( \begin{array}{cc} 1+p & \sqrt{p} + \sqrt{1-p}  \\ \sqrt{p} + \sqrt{1-p} & 2-p \end{array}\right)$. Similar to the noise free case, Bob measures the key in diagonal basis, and measures '0' with probability $\frac{1}{2}-\frac{\sqrt{p}+\sqrt{1-p}}{3}$. This implies, due to the noise in the quantum channel, Bob has a finite probability of measuring the key bit as '1', despite his measurement basis being same as Alice. One can also infer from Tables~\ref{tab_qstates_bp} and ~\ref{tab_qstates_ad}, that the errors generated in Bob's key can differ if the underlying noise in the quantum channel is different. In this work, we have provided a method to identify this difference by only looking at the keys generated at Alice and Bob's end, i.e., by analyzing the quantum bit error rate in their respective keys.   
In our protocol, to generate one key, Alice and Bob repeat the BB84 key generation process $16$ times, and compute the QBER as given in Eq.~\ref{eq_qber} for each key. Finally, we repeated the key generation process $200000$ times for each noise type, with random noise strength of the channels for each transfer, accumulating a list of $200000$ QBER values for both types of noises. \\
\vspace{5 mm}

\noindent\textbf{Data generation for classification of quantum noises:}
\begin{figure}[!hbt]
\centering
\begin{subfigure} {.4\textwidth}
\centering
        \includegraphics[width=1\linewidth]{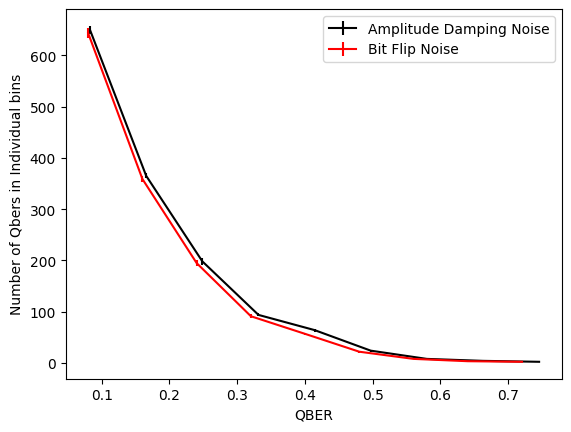}
        \caption{Histogram of QBERs}
    \label{theor_ad_bp}
    \end{subfigure}
    \begin{subfigure}{.4\textwidth}
    \centering
    \includegraphics[width=1\linewidth]{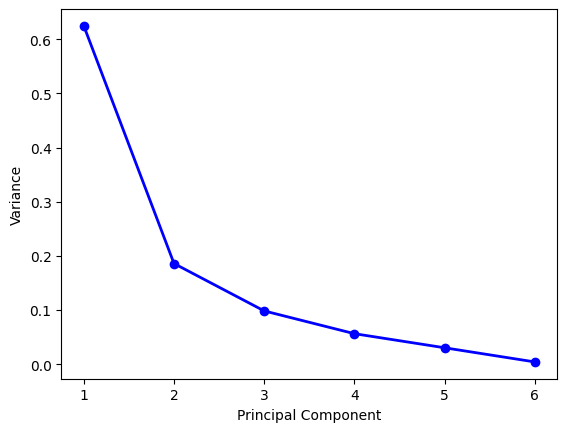}
    \caption{Scree plot}
    \label{theor_scree}
    \end{subfigure}
\caption{The average Histogram distribution of QBERs (only tip is shown) when two different noises (bit-flip and amplitude damping are applied to a quantum channel) is shown in Fig.~\ref{theor_ad_bp}. In Fig.~\ref{theor_scree}, the scree plot of the PCA of the dataset is presented. In this dataset, the first three elements can explain $91\%$ of the data-variance. To construct this dataset, we have taken $50$ blocks of the original data, and $4000$ QBERs per block.}
\label{theor_dataset}
\end{figure}

After accumulation of QBERs, we created individual histograms with $10$ bins for QBERs generated due to amplitude damping noise and bit-flip noise, by dividing our data into multiple blocks (say, $m$) for each noise, such that, each histogram contains $n= \frac{200,000}{m}$ QBER distributions. Fig.~\ref{theor_ad_bp} shows the average over all such histograms. It is clear from Fig.~\ref{theor_ad_bp}, that the distributions, on their own, are not properly distinguishable. However, as depicted by Tables~\ref{tab_qstates_bp} and \ref{tab_qstates_ad}, the retrieval probability of a certain key-bit by Bob differs under different noises. Thus, we planned to employ simple machine learning models to distinguish between different noise channels.
Next, we have chosen seven different features of the histogram distribution as the input to our classification model; mean, median, mode, standard deviation, skewness, kurtosis, and area under the curve, and generated a labeled dataset of $2 \times m$ data. As this dataset is small ($O(100)$), and assignment of QBERs to a particular block has no particular significance, to increase the number of datapoints in our dataset, with no loss of generality, we shuffled the original dataset randomly $100$ times, and repeated the histogram analyses, to finally create a larger dataset of $2m \times 100$ entries.\\
\vspace{5 mm}

\noindent{\textbf{Supervised Classification of noises:}} 
Next, the labeled dataset is divided into training set and test set with a 7:3 ratio. Next we perform a principal component analysis (PCA) \cite{Pearson01111901} to reduce the input features of the data . Fig.~\ref{theor_scree} shows the scree plot \cite{Cattell01041966} of the PCA of the dataset, which tells us that the first three principal components covers almost $91\%$ of the variance of the dataset. So, we choose the reduced features of the data to be $3$.

Finally we applied three different supervised classification models; K-Nearest Neighbor (KNN) \cite{KNN1,KNN2,KNN3}, Gaussian Naive Bayesian (GNB) \cite{GNB1,GNB2,GNB3}, and Support Vector Machine (SVM) \cite{SVM1,SVM2,SVM3,ker,rbf} on the training set. After training, we applied the models on the test dataset to compare the accuracy of classification.
We represent the accuracy of our classification method for training and test datasets in Table~\ref{tab_acc_theor}, for different values of $m$ and $n$.  
\begin{table}[!hbt]
    \centering
    \begin{tabular}{|c|c|c|c|c|}
    \hline
       \# Data-instances  & \# QBERs per histogram & ML model & Train Acc & Test Acc \\   
    \hline
     40000  & 1000 & KNN & $87.9 \%$ & $77 \%$\\
        &        & GNB & $71.9 \%$ & $72 \%$ \\
        &        & SVM & $77.1 \%$ & $77.2 \%$\\

    \hline
     20000  & 2000 & KNN  & $92 \%$ & $87 \%$\\
        &        & GNB & $58.5 \%$ & $56.9 \%$ \\
        &        & SVM  & $82.9 \%$ & $82.5 \%$\\

    \hline
     10000 & 4000 & KNN & $97 \%$ & $96 \%$\\
        &        & GNB & $67 \%$ & $64 \%$ \\
        &        & SVM & $92.5 \%$ & $91.5 \%$\\

     \hline
    \end{tabular}
    \vspace{2 mm}
    \caption{Accuracy comparison for quantum noise channel classification with classical machine learning models.  We have used the KNN model with $k=2$, and SVM with rbf kernel, degree $4$ for all the analyses.}
    \label{tab_acc_theor}
\end{table}

\begin{figure}[!hbt]
\centering
\begin{subfigure}{.5\textwidth}
  \centering
  \includegraphics[width=1\linewidth]{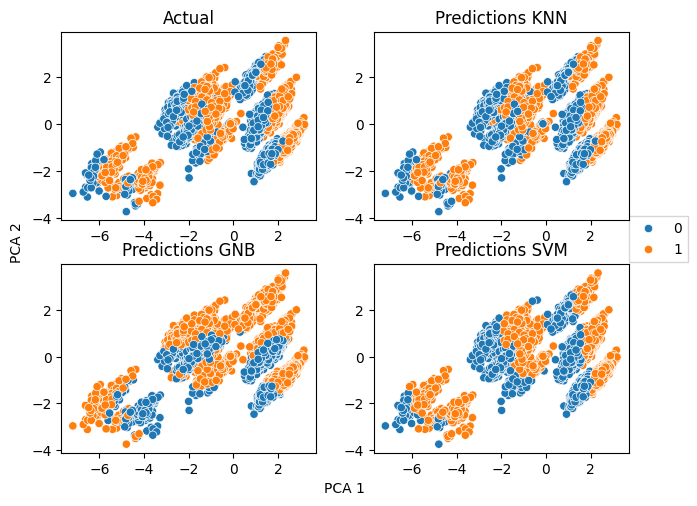}
  \caption{Training data}
  \label{theor_tr_3_class}
\end{subfigure}%
\begin{subfigure}{.5\textwidth}
  \centering
  \includegraphics[width=1\linewidth]{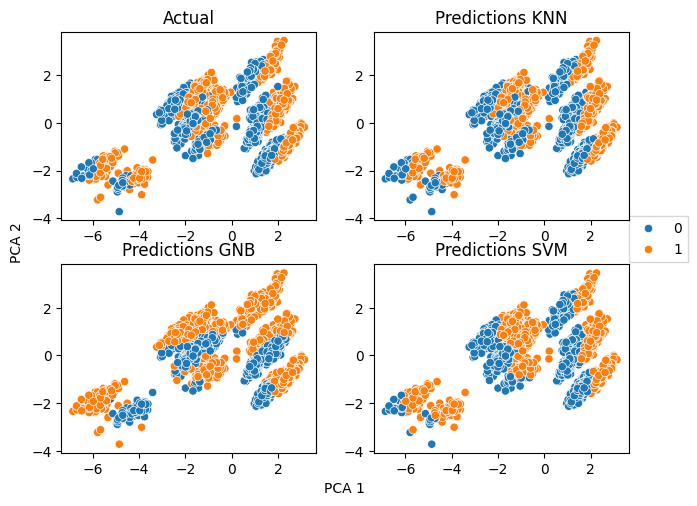}
  \caption{Test Data}
  \label{theor_test_3_class}
\end{subfigure}
\caption{Classification of quantum noise channels using QKD (BB84)protocol and trained K-Nearest Neighbors (KNN), Gaussian Naive Bayes (GNB), and Support Vector Machine (SVM) classifiers. $\text{PCA} 1$ and $\text{PCA} 2$ refers to the two features of the input dataset after Principal Component Analysis. The underlying QKD protocol used here is BB84. '0' and '1' respectively represent amplitude damping noise and bit-flip noise. Fig.~\ref{theor_tr_3_class} represents the classification of training data, and Fig.~\ref{theor_test_3_class} represents the classification of test data. Here $m$ and $n$ used is $50$ and $4000$ respectively.}
\label{theor_class}
\end{figure}

As can be seen from Table~\ref{tab_acc_theor}, the classification accuracy vary depending on the number of QBERs per histogram ($n$), number of total data in the classification dataset ($m$), as well as the machine learning model used. However, KNN with $k=2$ provides the best results for both training and test cases with $97 \%$ accuracy for training set and $96 \%$ accuracy for the test set, when the dataset contains a total of $5000$ instances per noise, but each QBER histogram distribution contains $4000$ entries. For the other two cases as well, we see KNN provides better accuracy than SVM (rbf kernel, degree=$4$) and GNB. To reason with this behavior, we take a look at Fig.~\ref{theor_class}. We can see from both Figs.~\ref{theor_tr_3_class} and \ref{theor_test_3_class}, the classification dataset is overlapping. As KNN works better with overlapping datasets, it provides a higher accuracy.  

\begin{figure}
    \centering
    \includegraphics[width=1\linewidth]{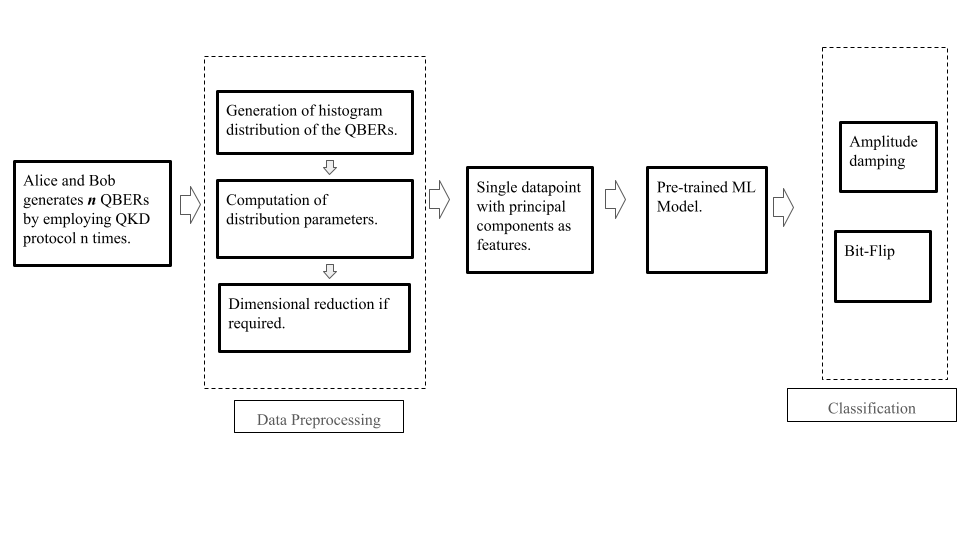}
    \caption{Workflow of the pretrained model for classifying two types of quantum noises from QBERs accumulated after a QKD scheme.}
    \label{workflow}
\end{figure}

Fig.~\ref{workflow} provides a general scheme for classifying two quantum noise channels using the trained model. 

\vspace{2 mm}
\noindent In this work, we have implemented the machine learning assisted noise classification protocol assuming two different scenarios. In one case, Alice and Bob are remotely located, and they communicate through a quantum channel, as explained in this section. In the other case, we assumed Alice and Bob have two different qubits of a noisy gate-based quantum computer at their disposal. In Sec.~\ref{qiskit_results} we provide our results for noise classification on gate-based quantum computers, and discuss our findings. 

\section{Noise classification in Gate-based Quantum computer}\label{qiskit_results}
\begin{figure}[!hbt]
\centering
\begin{subfigure}{.5\textwidth}
  \centering
  \includegraphics[width=.9\linewidth]{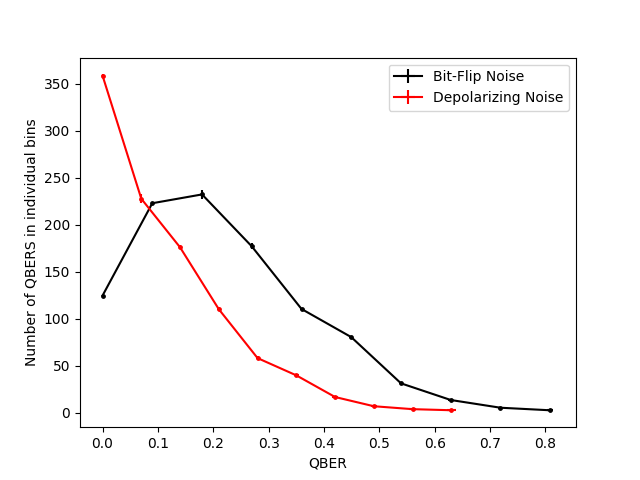}
  \caption{Distributions of QBERs in BB84.}
  \label{distri_bb84}
\end{subfigure}%
\begin{subfigure}{.5\textwidth}
  \centering
  \includegraphics[width=.9\linewidth]{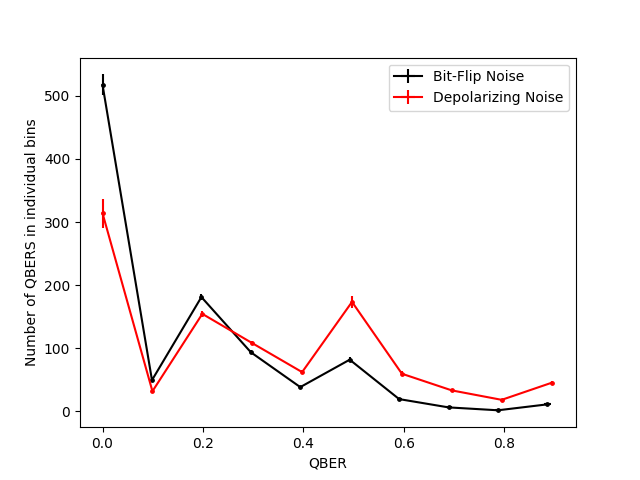}
  \caption{Distributions of QBERs in BBM92.}
  \label{distri_bbm92}
\end{subfigure}
\caption{Histogram distribution (only tip is shown) of QBERs when the quantum channel is affected by bit-flip and depolarizing noises with $10$ bins. Each plot-point is an average of $200$ histogram tips; Each histogram consists of $1000$ QBERs values. The distributions in Fig.~\ref{distri_bb84} represents QBERs originated from BB84, and  the distributions Fig.~\ref{distri_bbm92} represents QBERs originated from BBM92 QKD scheme.}
\label{distri}
\end{figure}
In this section, we show that our noise classification protocol can also be used to classify noises on a gate-based quantum computer. The broader outline of the protocol remains the similar as before, where a simplistic QKD protocol is simulated on a quantum computer under assumption of two different type of noises. Based on the outcome of Alice and Bob's measurements, a list of QBERs with $200,000$ values is recorded. Then, for each noise, the list of QBERs is divided in $m$ blocks, to generate a labeled dataset with $2m$ entries. This dataset has $7$ distribution parameters as features, and two output classes. Shuffling the distribution $100$ times, a larger dataset it prepared with $O(10000)$ entries. Further, principal component analysis (PCA) is performed on the dataset to reduce its features.  Finally we employ KNN, GNB, and SVM classifiers to train our classification model, after splitting the dataset into train and test data with a $7:3$ ratio.

\noindent To simulate a QKD protocol on a gate-based quantum computer under different noise channels, we have used the QISKIT software development kit offered by IBM Quantum \cite{qiskit2024}, and the in-built noise models therein. We have simulated both BB84 and BBM92 protocols, under assumptions of bit-flip and depolarizing noises. As before, we have not considered the readout error while simulating the QKD protocols.  
Fig.~\ref{distri} represents the average histogram distribution of QBERs for BB84 (Fig.~\ref{distri_bb84}) and BBM92 (Fig.~\ref{distri_bbm92}) protocols, for both types of noises considered. As can be seen from Fig.~\ref{distri_bb84}, the distributions are visually distinguishable in case of BB84, in comparison with Fig.~\ref{theor_ad_bp}. This difference stems from the difference of the source of the noise. In the case discussed in Sec.~\ref{method}, two remote parties Alice and Bob generate keys by communicating through a noisy quantum channel, whereas, in case of a gate-based quantum computer, the noises majorly come from the noises while applying a gate-operation. Further, Fig.~\ref{distri_bbm92} shows, in case of BBM92 simulated with qiskit under two different noise models, the distributions are distinct, although follow the same pattern. However, as is clear from Fig.~\ref{distri}, there are still significant overlap between the two noise types if we only consider the QBER distributions. Next, we have executed our protocol for feature extraction and classical machine learning algorithms for distinguishing the quantum noises impacting a gate-based quantum computer. 
\begin{figure}[!hbt]
\centering
\begin{subfigure}{.5\textwidth}
  \centering
  \includegraphics[width=1\linewidth]{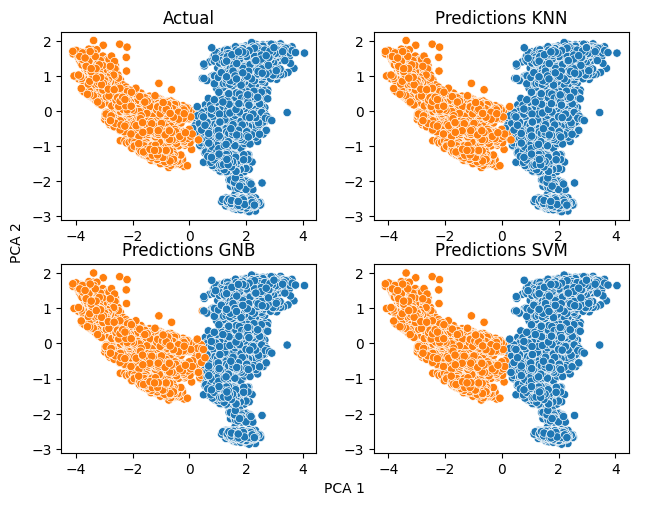}
  \caption{Training data}
  \label{BB84_tr_3_class}
\end{subfigure}%
\begin{subfigure}{.5\textwidth}
  \centering
  \includegraphics[width=1.25\linewidth]{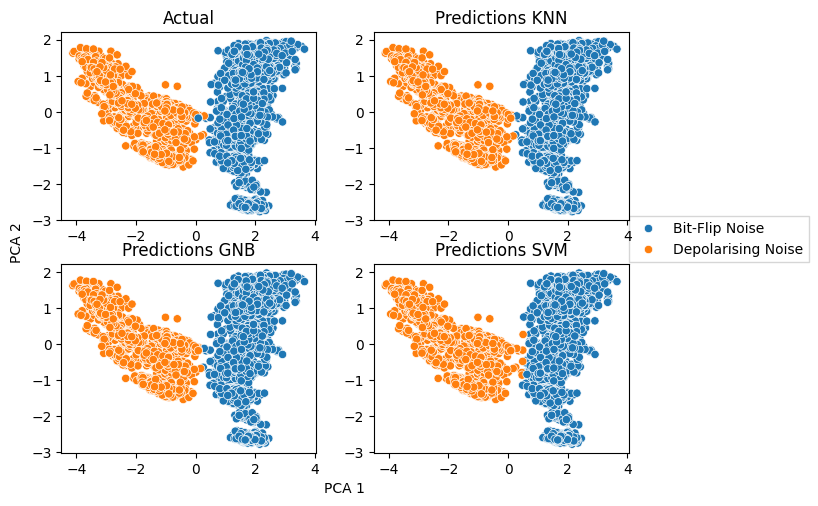}
  \caption{Test Data}
  \label{BB84_test_3_class}
\end{subfigure}
\caption{Classification of test-set data with trained K-Nearest Neighbors (KNN), Gaussian Naive Bayes (GNB), and linear Support Vector Machine (SVM) classifiers. $\text{PCA} 1$ and $\text{PCA} 2$ refers to the two features of the input dataset after Principal Component Analysis. The underlying QKD protocol used here is BB84.}
\label{BB84_classification}
\end{figure}

\begin{figure}[!hbt]
\centering
\begin{subfigure}{.5\textwidth}
  \centering
  \includegraphics[width=1\linewidth]{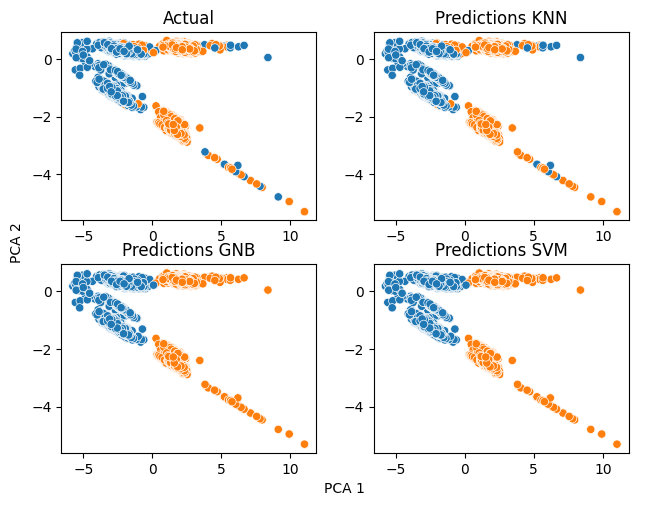}
  \caption{Training data}
  \label{BBM92_tr_3_class}
\end{subfigure}%
\begin{subfigure}{.5\textwidth}
  \centering
  \includegraphics[width=1.25\linewidth]{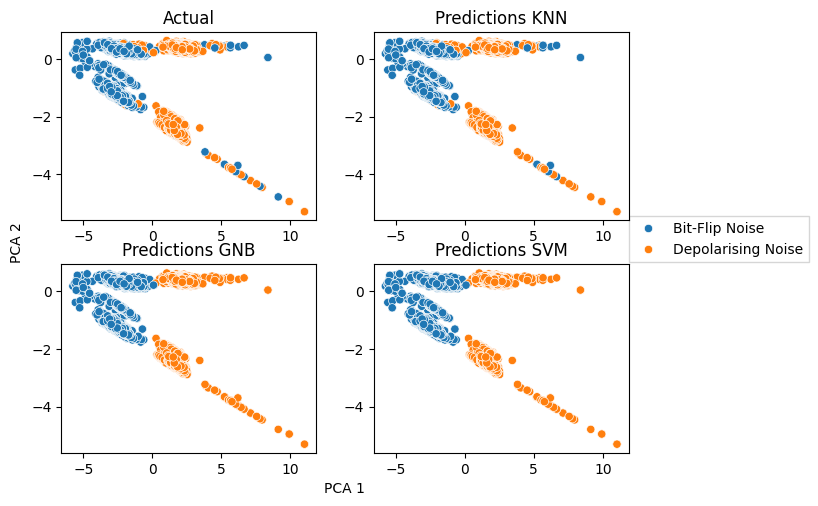}
  \caption{Test Data}
  \label{BBM92_test_3_class}
\end{subfigure}
\caption{Classification of test-set data with trained K-Nearest Neighbors (KNN), Gaussian Naive Bayes (GNB), and linear Support Vector Machine (SVM) classifiers. $\text{PCA} 1$ and $\text{PCA} 2$ refers to the two features of the input dataset after Principal Component Analysis. The underlying QKD protocol used here is BBM92.}
\label{BBM92_classification}
\end{figure}

As mentioned earlier, we have implemented our protocol with two different QKD protocols. For simulations with BB84, we have generated and used a labeled dataset of $20,000$ entries, and initially $7$ features. After PCA, we have reduced the number of features to $3$ as for this dataset, the first three principal components cover $93 \%$ of the variations. Finally, we trained KNN (k=2), Linear SVM, and GNB classifiers with a training set of $14000$ data. Finally, we tested our models on a test dataset of  $6000$ data. For the dataset generated BBM92 protocol, there are $40,0000$ entries, and $7$ features. However, we have used $2$ principal components covering $99 \%$ of the data variance for training and testing our machine learning models.

\begin{table}[!hbt]
    \centering
    \begin{tabular}{|c|c|c|c|c|c|}
    \hline
      QKD & \# Data-instances  & \# QBERs per histogram & ML model & Train Acc & Test Acc \\   
    \hline
      BB84 & 20000  & 1000 & KNN & $99.9 \%$ & $99.9 \%$\\
       &  &        & GNB & $99.9 \%$ & $99.8 \%$ \\
        &   &      & SVM & $99.8 \%$ & $100 \%$\\

    \hline
     BBM92 & 40000  & 2000 & KNN & $99.8 \%$ & $99.7 \%$\\
       &  &        & GNB & $99.7 \%$ & $99.6 \%$ \\
        &   &      & SVM & $99.7 \%$ & $99.6 \%$\\

    \hline
    \end{tabular}
    \vspace{2 mm}
    \caption{Accuracy comparison for quantum noise channel classification with classical machine learning models.  We have used the KNN model with $k=2$, and SVM with linear kernel for all the analyses.}
    \label{tab_acc_qiskit}
\end{table}
Fig.s~\ref{BB84_classification} and \ref{BBM92_classification} portrays our results on the training and test datasets with BB84 as the underlying QKD protocol. Further, we tabulate the accuracy of our protocol with different machine learning classifiers as well as two different QKD protocols in Table~\ref{tab_acc_qiskit}. For this case, our classification methods works really well with $99 \%$ accuracy across datasets, classifiers, and QKD protocols. From Fig.s~\ref{BB84_classification} and \ref{BBM92_classification} we attribute this high accuracy to the non-overlapping classes of the dataset.

\section{Conclusion and Future directions}\label{conc} 
In this work, we have proposed a novel method to classify quantum noise channels with only the quantum bit error rates of simplistic QKD protocols using supervised machine learning. The contribution of our work ranges from building up the model from scratch to obtaining a high accuracy, including identifying the input features required to train for such classification tasks. We have shown the efficiency of our classification protocol in two different cases: we theoretically simulated BB84 protocol assuming amplitude damping and bit-flip channels between remotely located Alice and Bob, and have shown that the noisy channels can be classified with $96 \%$  accuracy. Subsequently, we have shown our protocol is able to distinguish two different noises, depolarizing and bit-flip, afflicting a gate-based quantum computer with $99 \%$ accuracy, by using two different QKD schemes, BB84 and BBM92.  

Alongwith our primary findings, we have shown, under the assumption of remotely located Alice and Bob, the histogram distribution of quantum bit error rates (QBERs) originating due to amplitude damping and bit-flip noises have overlapping characteristics, leading to classification with high accuracy possible only with K-nearest neighbor classifier. Further, we have shown that the classification accuracy increases if we assign more data to analyze the histogram features, instead of having a larger dataset. Subsequently, in case of gate-based quantum computer, where the noises attacks only the gates used, the distributions have non-overlapping features, leading to even a $100 \%$ accuracy with liner SVM. We have also shown that this observation is independent of the choice of QKD protocol. 

In future, we plan to extend our study to mixed noise channels, as well as study the performance of our protocol using real data.  

\bibliographystyle{unsrt}

\end{document}